# AMIC: An Adaptive Information Theoretic Method to Identify Multi-Scale Temporal Correlations in Big Time Series Data

Nguyen Ho*, Huy Vo†‡, Mai Vu§, Torben Bach Pedersen*
*Department of Computer Science, Aalborg University, Denmark
†Center for Urban Science and Progress, New York University, New York, USA
‡Department of Computer Science, the City College of New York, New York, USA
§Department of Electrical & Computer Engineering, Tufts University, Medford, MA, USA
Email: ntth@cs.aau.dk, huy.vo@nyu.edu, mai.vu@tufts.edu, tbp@cs.aau.dk

**Abstract**—Recent development in computing, sensing and crowd-sourced data have resulted in an explosion in the availability of quantitative information. The possibilities of analyzing this so-called Big Data to inform research and the decision-making process are virtually endless. In general, analyses have to be done across multiple data sets in order to bring out the most value of Big Data. A first important step is to identify temporal correlations between data sets. Given the characteristics of Big Data in terms of volume and velocity, techniques that identify correlations not only need to be fast and scalable, but also need to help users in ordering the correlations across temporal scales so that they can focus on important relationships. In this paper, we present AMIC (Adaptive Mutual Information-based Correlation), a method based on mutual information to identify correlations at multiple temporal scales in large time series. Discovered correlations are suggested to users in an order based on the strength of the relationships. Our method supports an adaptive streaming technique that minimizes duplicated computation and is implemented on top of Apache Spark for scalability. We also provide a comprehensive evaluation on the effectiveness and the scalability of AMIC using both synthetic and real-world data sets.

✦

## 1 INTRODUCTION

**Motivation** The recent notable development of the so-called Big Data brings both excitement and confusion to the research and industry communities. The availability of massive, heterogeneous and rich data sets promises to contain interesting patterns and trends that once unlocked can enable actionable intelligence and evidence-based decision making. However, the tremendous amount of data generated with high velocity and complex formats introduces significant challenges in terms of computational reliability and efficiency. The requirement of having scalable methods to efficiently analyze large data sets becomes crucial to capture and unveil insightful information in real-time.

Indeed, Big Data is worthless in the absence of such methods. Only through analyzing multiple and crossed data sets, the true value of Big Data can be harnessed. One of the first steps towards creating new value in Big Data applications is the discovering of correlations among heterogeneous and cross-domain data sets. As data originally reside in individual silos, each of them may serve for a specific and/or limited purpose. However, their combination can, and often does, offer new insights into important problems. Particularly in data exploration, data correlation can result in the identification of individual events and phenomena, as well as the creation of profiles to track these activities even in real-time. Data correlation is also useful in constructing and validating behavioral proxies. For example, demonstrating that the traffic speed is well-correlated with the number of taxi pickups through historical data sources for which the latter are known directly (e.g., through the NYC Taxi & Limousine Commission) would allow accurate measurement of traffic speed in real-time. More broadly, finding correlation among data sets will allow policy makers to better understand cities, thus, providing better operations and better planning to citizens. In the finance sector, data correlation can help businesses make better investing decisions through forecasting the price movement of related stocks, or predicting purchasing behavior of consumers. In areas such as health-care, through identifying the association between variables, data correlation provides the first clues to uncover root causes of phenomena of interest (e.g., a disease or a health condition). Although not implying causation, data correlation is one of the three criteria for establishing a causal relationship between two variables, as argued by Agresti et al. in [1], thus discovering data correlations in heterogeneous data sources can be very useful when building causal models.

**Challenges and limitations of current approaches** Despite the promising use of correlations in the Big Data era, finding correlations in Big Data corpuses is a hard problem. Not only does the abundance of data make it impractical to manually determine the correlation between them, their usually large temporal coverage is also a challenge in identifying periods of interest (e.g. an event) at different data resolutions. Considering NYC Open Data [2], with more than 1,500 data sets that have been published and updated since 2009, it would take an immense amount of effort for an analyst just to select a data set that is well-correlated with another data set of interest and combine them together. The challenge still persists even when correlated data sets have been identified. For instance, when a policy adviser wants to study the impact of taxi cabs on 311 complaints, s/he would like to know, e.g., the particular day or week when the two data sets have the highest correlation (if any). It is not only necessary to know whether two data sets are well-correlated, but also when the correlations are the strongest.

Moreover, with Big Data the real challenge does not come from its big *volume* alone, but more from its *variety* and *velocity* properties where structured and unstructured data are continuously and rapidly generated from large number of sources. Methods to discover correlations in continuous high-speed and complex time series streams have to be scalable and efficient in order to deal with billions of data items quickly. The methods also have to be robust enough to deal with noisy data which very likely contain complex and non-linear relations. This goes with the requirement of having a correlation measure that can capture different types of relations (e.g., linear vs. non-linear, functional vs. non-functional), and has the ability to rank and establish orders of discovered relationships.

When searching for correlations among heterogeneous data sources, it is important to define criteria to determine when an association is significant. For example, defining a threshold where a combination above the threshold will be significant and vice versa. Setting the threshold too low can result in many uninteresting combinations, whereas setting the threshold too high can potentially miss interesting associations. Having an appropriate method to set this threshold so that correlations can be discovered even when data characteristics and their relationships are unknown is an important challenge to be addressed. Moreover, as real-world data can have patterns where correlations can appear strong at one point but weak or completely disappear at others, it is also important to have an approach that can search for correlations regardless of data resolution.

Although significant work has been done in finding correlations between data sets, relatively little investigation has been made with adaptive temporal scales. Most of the work in correlation analysis assume a fixed time scale for each data set. For example, finding correlation between taxis and weather will only return well-correlated periods either in hours, in days, or in weeks. However, in practice, it is common to see strong correlations between two data sets at multiple temporal scales ranging from hours (e.g. during rain showers) to weeks (e.g. during a storm). Additionally, when multiple correlated periods are found, they are often presented in the same way regardless of their significance. Users then have to rank them manually for their analysis.

**Contributions** In this paper, we present AMIC, a scalable and efficient framework for identifying multi-scale temporal correlations in big data sets, addressing the *volume*, *velocity* and *variety* challenges present in Big Data context. Particularly:

- We use mutual information (MI) as a correlation measure to capture and quantify the co-dependence between variables, allowing us to discover different types of relations potentially present in Big Data context (e.g., linear and non-linear, functional and non-functional). By quantifying the dependency, our method has the ability to rank the significance of discovered correlations automatically. Moreover, as mutual information is a solid statistical measure of co-dependence that works for different types of data (e.g., text, numbers), AMIC helps overcome the *variety* challenge of Big Data.
- AMIC is built based on an adaptive sliding window technique, thus enabling us to discover not only the correlated data sets, but also when the correlations occur, and when they are strongest. We also propose an optimized search method to address two Big Data challenges:

  – To handle large data *volume*, a layering and partitioning mechanism is used together with an optimized computation technique. Data partitioning provides the ability to handle billions of data items while the layering approach automates filtering non-useful data and adaptively changes resolution, thus supporting multi-scale temporal resolutions. The optimization technique helps minimize duplicated computation across streaming windows, and thus provides an efficient search to meet timeliness requirement.

  – To address the *velocity* challenge, the method is designed in streaming fashion to support real-time computation. The (nearly) real-time response is achieved thanked to the distributed recursive parallelizing mechanism, and the optimized MI computation technique. Moreover, the search method returns window-based correlations which enables us to trace the time intervals of correlation windows. This allows us to suggest *highly interesting* temporal periods to users for further investigation. For example, window-based correlation between taxi trip and wind speed allows users to trace back extreme weather events happened in NYC.

- We propose three different methods to define the correlation threshold, and provide the justification about their usage in different contexts. The proposed methods assist users in setting an appropriate threshold even when no prior knowledge of data characteristics and their relationships are available.
- We propose the formalization of positive and negative associations to justify the nature of discovered correlations in extracted windows.
- To meet the scalability requirement, we design and implement a recursive parallel computation mechanism on top of the Apache Spark platform. The mechanism divides large data series into multiple smaller series, and recursively distributes them among the worker nodes within the Spark clusters to exploit the computational availability of the Big Data platform.
- We perform a comprehensive evaluation of our techniques on both synthetic data sets and real-world data sets (including the NYC open data, and the energy production data). Synthetic data sets help evaluate the correctness and the scalability, whereas real-world data sets demonstrate the applicability and relevance of our proposed method.

## 2 RELATED WORK

In this section, we review related work on the correlation metrics, the correlation discovery techniques and the use of mutual information in data analysis.

*The correlation metrics:* Traditional statistical metrics such as covariance, correlation coefficients (e.g., Pearson, Spearman) have been used extensively in the literature, for example in [3]–[5], to identify correlations among data. However these metrics are limited to linear and monotonic (i.e., strictly increasing or strictly decreasing) dependencies. Zhang et al. in [6] use *correlation coefficient* (*corr*) to analyze correlations in spatio-temporal data sets. In order to reduce computational cost, these authors exploit spatial autocorrelation among spatial neighboring time series and group similar time series into cones before computing correlations across data sets. Although the work uses *corr*, and thus only works with linear dependencies (unlike our work), the idea is interesting and could be exploited in a future extension of our work for both temporal and spatial dimensions. Other work exploits spatio-temporal correlations to serve different purposes, e.g., to control and optimize Wireless Sensor Network [7], video coding [8], anomaly detection [9], and data compression [10], [11].

Recent work such as [12]–[18] attempts to approach the problem from a high level. Sarma et al. [12] propose a framework to find related tables in a large corpus of databases, using the concept of *relatedness* to capture different kinds of related-relation between data tables. In [13], Pochampally et al. propose to model correlations between different data sources using *joint precision* (portion of correct outputs over entire outputs) and *joint recall* (portion of all correct triples that are output by all sources) as indicators. In comparison, the work in [14] relies on history and schema of data sets to map and link them together. In [15], Roy et al. use the concept of *intervention* (i.e, changes in the values of inputs affect the outputs) to look for causal explanation for the answers of SQL queries. Yang et al. [16] use a *residue* metric that measures the difference between the actual and expected value of an object to capture objects correlation in large data sets. Sousa et al. [17] propose a fast method to find correlations among attributes in databases, using the concept of *intrinsic dimension* (a small subset of dimensions that can represent entire data) to reduce redundant attributes. Middelfart et al. [18] propose a bitmap-based approach to discover *schema level* relationships in a multidimensional data cube. These works differ from our approach as we aim to look for not only relationships between data sets but also the time windows where the data are well-correlated.

In a recent work [19], Chirigati et al. propose a topology-based framework to identify relationships between spatio-temporal data sets. The notion of topological features, whose *interestingness* is captured by critical points (maximum and minimum), is defined to represent the data sets and identify relationships between them. The technique is able to find both regular relationships and relationships at extreme events. Our work is aimed at achieving similar goals, and at the same time also picks out the exact time windows where the correlations occur, and thus can be used in conjunction with their work to have a more comprehensive technique and improve the data exploration process.

*The correlation discovery techniques:* Using window-based techniques to find correlations is not a new idea. For example, the work in [20], [21] also uses this technique to search for correlations. However, they adopt different correlation metrics, e.g., Schulz et al. [20] use Spearman coefficient, while Cole et al. [21] use sketches. By using MI as a measure and proposing robust methods to define the correlation threshold, our work has the advantage of being able to discover both simple and complex relations (more in Section 3), and thus fits better into Big Data context. In addition, we propose a divide-and-conquer approach to avoid the common combinatorial explosion barrier of correlating time-series data from large and varying data sources, as well as introduce an optimized and scalable computation framework built on top of a leading Big Data platform (Apache Spark), and thus provide a more up-to-date solution to the problem of correlation finding in the Big Data era.

In the context of data streaming, a recent work [22] of Keller et al. proposes the MISE algorithm to estimate MI for time series streams. What is different from our work is that we use a top down approach to partition the data and adaptively change its resolution, and thus help minimize the search space, while at the same time still ensure the capture of significant correlations. We introduce the concept of *influenced region* to keep track of changes in the data, a technique not used in [22]. Other work such as [23] instead uses Discrete Fourier Transform to detect local correlations in streaming data. Their focus, however, is on the delay and linear local correlation.

In another recent work [24], Bermudez et al. use a sliding window technique with two metrics, Pearson coefficient and entropy, to discover correlations in spatio-temporal big data. This work, however, differs from our work in several significant ways. Instead of using just a single-size window as in [24], our method uses a multi-layer sliding window technique, and thus, can discover correlations at multiple temporal scales. We further design an optimized algorithm using specific data structures to minimize redundant computation. Our framework also offers the ability to rank the relationships, so that discovered correlations can be ordered according to their strength and significance.

*The use of mutual information:* The MI measure has been broadly used in numerous domains to achieve different goals, e.g., feature selection [25], [26], clustering and mining [27], [28], image alignment and registration [29], network inference and construction [30], [31], dependency discovery in glucose measurements [32], or spatial temporal dynamic of the magnetosphere [33]. In the context of Big Data, the use of MI is relatively new. A recent work of Su et al. [34] proposes a framework and a set of algorithms to analyze relationships of massive scientific datasets, in which MI is one of the metrics to measure correlations. However, they only consider overall correlation and focus on data indexing to efficiently compute correlation in parallel and distributed setting.

The basic structure of AMIC has been initially investigated in our previous work [35]. In the present paper, on top of a broad update of literature, we make the following extensions to [35]: we propose three different theoretical methods to set the correlation threshold, and justify their usage in different contexts. We re-design AMIC using these new threshold metrics. We propose the formalization of positive and negative associations to justify the correlations in extracted windows. We also improve AMIC performance by introducing a new data structure, the *influenced marginal region*, to keep track of changes in incremental computation. We also design and implement a distributed recursive parallel search method to address the *velocity* challenge of Big Data and improve the scalability of AMIC. We make an extensive evaluation of AMIC using both synthetic and real world data sets, as well as perform a large-scale stress test and scalability test using Spark clusters. These extensions further formalize the AMIC framework and strengthen its effectiveness in the current Big Data landscape.

## 3 BACKGROUND

### 3.1 Mutual Information: Definition and Properties

In information theory, MI has been used as a measure of mutual dependency between variables. It quantifies the amount of information obtained about one variable through the knowledge of other variables [36]. The mutual information, $I(X, Y)$, between two random variables $X$ and $Y$ specifies how much knowledge about $X$ is gained through $Y$, or how much uncertainty of $X$ is reduced by knowing $Y$ and vice versa. MI is a function of the probability distribution, i.e., the probability density function (pdf) for continuous variables and the probability mass function (pmf) for discrete variables, and can account for both linear and non-linear relationships. Eq. 1 defines the MI of two discrete random variables $X$ and $Y$:

$$I(X,Y) = \sum_{y \in \{Y\}} \sum_{x \in \{X\}} p(x,y) \log \frac{p(x,y)}{p(x)p(y)} \quad (1)$$

where $p(x,y)$ is the joint probability of $(X, Y)$, and $p(x)$, $p(y)$ are the marginal probabilities of $X$ and $Y$, respectively.

From Eq. 1, $p(x, y)$ measures the probability that $X$ and $Y$ are observed together, while $p(x)$ and $p(y)$ are the probabilities that $X$ and $Y$ occur separately. The fraction $\log(p(x,y)/(p(x)p(y)))$ determines the magnitude of joint occurrence over the individual realization of the variables. The larger this magnitude, the more likely these two random variables occur together and thus, more likely they are dependent on each other. Intuitively, if the two variables are statistically independent, their MI is zero, meaning that knowledge of one variable does not reveal anything about the other. On the other hand, if the two variables are statistically dependent, their MI will be greater than zero, and attains a larger value as the dependency between the two variables becomes stronger.

Mutual information has several properties making it advantageous when evaluating correlations over other measures such as covariance or correlation coefficient [37]. First, mutual information is equal to zero if and only if the considered variables are statistically independent, otherwise positive if they hold any kind of dependency (e.g. functional, non-functional [38]). This property makes MI a versatile measure to capture correlations and ideal for noisy data sets which exhibit high degree of bias and abnormality, causing their relationships often arbitrary and non-linear. Second, mutual information is invariant under 1-1 transformations, i.e, $I_{XY} = I_{UV}$ if $u = u(x)$ and $v = v(y)$. This property says that under the transformation, if $X$ and $Y$ maintain their distributions, their MI is preserved. This characteristic is ideal in processing spatio-temporal data sets which are often collected beforehand under different resolutions. For example, in our reference case study, taxi data is collected every minute while traffic speed is recorded each hour.

### 3.2 Estimating Mutual Information

Although MI is a powerful measure in discovering relationships among data sets, it is challenging to apply in practice due to difficulties when estimating probability distributions. Among several estimation methods [39] (e.g., histogram, kernel density estimation), we choose a popular non-parametric method proposed by Kraskov et. al. [40], hereafter called the *KSG* method, to approximate MI because of the following reasons: (1) This method outperforms other estimators in terms of computational efficiency, accuracy and is especially suitable for long and chaotic time series [41]; (2) The method uses k-nearest neighbor approximation and thus is data efficient (i.e., it does not require very large samples), adaptive and has minimal bias [40]. These reasons make the method particularly suitable to study spatio-temporal data where dependence between variables might only occurs at specific times or locations.

*KSG Mutual Information Estimator* The main idea of *KSG* estimator is that rather than directly computing the joint and marginal probability distributions of considered variables, it estimates the densities of data points in neighborhoods [40]. For each data point, it first searches for $k$ nearest neighbor clusters ($k$ is a pre-defined parameter) and computes distance $d$ to the $k^{\text{th}}$-neighbor. Then, the population density within distance $d$ is estimated by counting the number of data points that fall inside $d$. This leads to the computation of MI between $X$ and $Y$ as [40]:

$$I(X,\ Y) = \psi(k) - 1/k - \langle \psi(n_x) + \psi(n_y) \rangle + \psi(N) \quad (2)$$

where $\psi$ is the digamma function, $k$ is the number of nearest neighbors, $(n_x,\ n_y)$ is the number of marginal data points in each dimension falling within the distance $d$, $N$ is the total number of data points and $\langle \cdot \rangle$ is the average function.

The intuition behind this estimator is, as MI aims to seek for knowledge of $X$ based on $Y$ (or vice versa), it looks into $Y$'s neighborhood and checks if nearby values $y_i$ result in closely related values $x_i$. It means that, for a specific data point, if its neighborhood in the $(X, Y)$ space corresponds to similar data, then knowing $Y$ helps predicting $X$ and vice versa, implying a high MI between X and Y. This concept is illustrated in Fig. 1, plotting taxi trips (counted by hour) versus wind speed (averaged by hour) in NYC during two different periods: the samples in blue are recorded when the Sandy hurricane was approaching NYC from $29^{th}$ Oct 2012 to $30^{th}$ Oct 2012, the samples in red are data recorded during normal days. As we can see, if one looks into the blue neighborhood, one can find high values of wind associated with abnormally low values of taxi, while the red neighborhood records more diverse data with low wind values associated with a wider range of taxi values. As can be seen, the blue neighborhood shows a clearer pattern, and as we will see, yields higher MI.

*Choosing the value of $k$:* The *KSG* method requires a free parameter $k$, i.e., the number of nearest neighbors to be searched for each data point. A study of different MI estimators in [41] shows that the k-nearest neighbor (kNN) estimator is the most stable method, and is least affected by the method-specific parameter, i.e., $k$. This motivates our choice to use the kNN estimator. The study in [41] also suggests that $k \leq 8$ provides a highly accurate estimation. Note also that larger $k$ will result in expensive computation. In general, there is no theoretical basis for selecting an optimal value for $k$. In practice, an empirical method is often used to determine an appropriate $k$ value. In the present work, our empirical method is based on a tuning process to find $k$, using our real data sets. Specifically, we use different values of $k$, e.g., $k$ is from 1 to 20, to compute MI and compare the variance of MI under these different $k$. The $k$ value that gives the least variance is selected. In Section 5, we discuss the result of this tuning process (Figure 7).

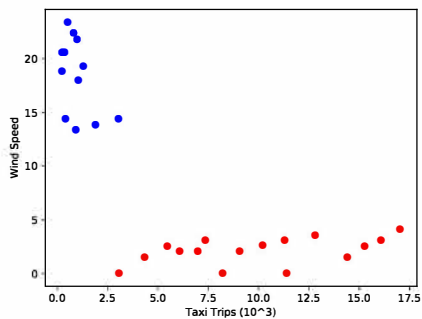

Fig. 1: Taxi Trips vs. Wind Speed during normal days (red) and during hurricane (blue)

## 4 HIERARCHICAL LAYERED CORRELATION SEARCH USING ADAPTIVE SLIDING WINDOWS

For real-world data, correlations might be present at different temporal periods. We design a hierarchical layered search algorithm based on sliding windows, and perform the search in a top-down fashion. Each layer in the search hierarchy works on one temporal scale, and the computation in each layer is performed incrementally. In the following, we provide the problem definition in Section 4.1, and describe the hierarchical search architecture in Section 4.2. We discuss the Spark parallel implementation of the algorithm in Section 4.3. When searching for correlations, it is also important to know when an interesting correlation has been found, and the nature of the discovered correlation, i.e., negative or positive. These topics will be discussed in Section 4.4 (correlation threshold methods) and Section 4.5 (positive/negative correlations).

### 4.1 Using Mutual Information to Measure Correlations in Time Series

Let $X = \{x_t\}_{t=1}^{N}$ and $Y = \{y_t\}_{t=1}^{N}$ be two finite time series of equal length $N$. The joint time series between them is $(X, Y) = \{(x_t, y_t)\}_{t=1}^{N}$. Since $X$ and $Y$ are sampled in time, they are either discrete or discretized. The frequencies of $(x, y)$ combinations can be computed by counting the number of times each combination occurs in the data, and then used to estimate MI value. Through estimating the MI of $(X, Y)$, we want to seek the solutions for the following problems:

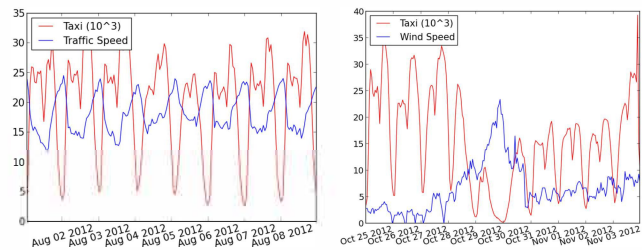

(a) Taxi Trips and Traffic Speed  (b) Taxi Trips and Wind Speed

Fig. 2: Different patterns in time series

*Problem 1.* Determine if variables $X$ and $Y$ are overall correlated, and if so, the strength of their relationship.

*Problem 2.* If overall correlation does not hold for $X$ and $Y$, then search for time windows $w_{ij} = [t_i, t_j]$, $1 \leq i < j \leq N$, where $X$ and $Y$ are highly correlated.

A positive MI value between variables $X$ and $Y$ over their entire data series indicates they are correlated in general. Thus *Problem 1* can be answered by computing MI for the entire $(X, Y)$ series. The relationship strength can be determined using MI magnitude: the larger this magnitude, the stronger the relationship. As an example, consider two variables: number of taxi trips $(X)$ and traffic speed $(Y)$, shown in Fig. 2a (the data are taken from NYC open data sets [2]). We might observe that whenever the number of taxi trips $x_i$ is high, the traffic speed $y_i$ is low, or vice versa, implying a pattern that high number of taxi trips might slow the traffic. If we ever wonder whether these two variables have any correlations, we can take their entire time series data and compute the MI. If it is positive, there exists the dependency between them.

On the other hand, consider taxi trips and wind speed in Fig. 2b, there is not a clear pattern between these two except for the period from Oct $29^{th}$ to Oct $30^{th}$ where we observe a significant drop in taxi trips associated with extremely high wind speed. This period is when hurricane Sandy approached NYC causing abnormally high wind. Thus the correlation between these two might not exist during regular days, but only in extreme events. In this case, MI of the pair (taxi trips, wind speed) might be very low during regular days but significantly high in the time window [Oct $29^{th}$, Oct $30^{th}$]. Looking for such time windows throughout the time series will answer *Problem 2*.

### 4.2 Layered Search with Adaptive Sliding Windows

Our goal is to efficiently search for multi-scale temporal correlations over time series data. One possible (*naive*) approach is to compute the MI for all possible temporal periods. This approach however is computationally expensive because of the combinatorial explosion of possible windows. To be more efficient, we design a hierarchical algorithm composed of multiple layers, where each layer is responsible for one temporal granularity. Windows at different layers will have different sizes, while in the same layer share the same size. At each layer, same size windows are slided over the time series, iteratively filtering data partitions where interesting correlations might be present. Initially, the search starts with the coarsest granularity, i.e., the largest window size. Data partitions that do not contain interesting correlations in that layer will be passed to lower layers, to be searched with finer granularity. The advantages of this multi-layer approach are: (1) multi-scale temporal correlations can be uncovered using different window sizes, (2) computation cost is minimized because the lower layers only work on data filtered out by the upper layers. Moreover, the computation within the same layer is performed incrementally, thus eliminating redundant computation. In the following, we provide the definitions used in our algorithm, and detail each computation step in the subsequent sections.

*Definition 1.* Given two time series $X = \{x_t\}_{t=1}^N$ and $Y = \{y_t\}_{t=1}^N$, a *time window* $w_{X,Y}$ of $(X, Y)$ is a sequence of timestamped data points $(x_t, y_t)$ collected over a continuous time interval and sorted in chronological order.

*Definition 2. Window granularity* is a temporal unit representing the time scale of a window. For example, a window stores data collected in *hours*, *days*, *weeks*, and *months* will have *hour*, *day*, *week*, *month* granularities, respectively.

*Definition 3. Window size* is the length measured by the number of data points contained in the window. A window must have proper size, i.e., contain enough samples, in order to report any significant correlations.

*Definition 4. Sliding step* is the moving step by which the window will be shifted from its current time window to the next time window.

*Definition 5. Threshold* $\sigma$ is a non-negative real number representing the minimal level of correlation required for significance. A window that has its correlation value $\geq \sigma$ is said to contain significant dependency. In Section 4.4, we propose three different ways to define this threshold, and discuss their usage in relevant contexts.

### 4.2.1 Top-down Filtering with Adaptive Window Size

We allow user to define the *maximum* ($g_{max}$) and the *minimum* ($g_{min}$) granularity s/he wants to operate on the data, as well as the *sliding step* by which the window will be shifted. Starting with the largest granularity, data are partitioned into windows of the same size. Two consecutive windows can be disjoint or overlapping depending on whether significant correlation exists in one of them (details are described in the next section). Using *KSG* estimator, the MI value is computed for each window and is compared against the threshold $\sigma$. Windows that have significant correlations (i.e., $\geq \sigma$) are selected. After the first scan, a set of *selected* windows is returned. The *filtered-out* windows (called *left-out* data) are those that do not satisfy the defined threshold and will be used for the next scan with finer granularity. The search stops when the granularity reaches the *minimum*, or the search procedure has scanned all data, and all data are included in the *selected* windows. With this layered top-down approach, user has the flexibility to look for dependencies at different time scales. To test if variables of interest are overall correlated, user can set the *maximum* granularity covering the entire data series, while decreasing the granularity to a finer scale helps uncover correlations in separate time periods. For example, when searching for correlations between taxi and weather data, user can start with *year* granularity, then reduce the granularity to *month*, *week*, *day*, and *hour* if significant correlations are not found.

### 4.2.2 Sliding Windows with Filtering

This step works at each granularity layer. At each layer, the procedure *search_windows* moves windows of same size along the data series and select those that satisfy the defined correlation threshold. We illustrate this movement in Fig. 3.

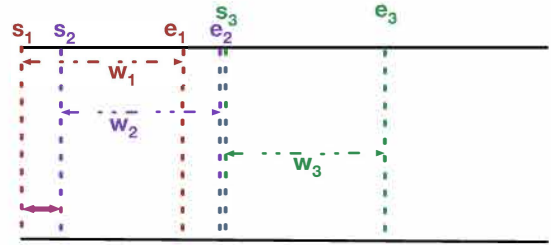

Fig. 3: Sliding windows search with filtering

Consider the pair of variables $(X, Y)$, each window $w_i$ is identified by the start index $s_i$ and the end index $e_i$ (indicating the first and the last data point of $w_i$). Since each data point is associated with a timestamp, the start and end index also indicate the window's start and end time. We use the data structures *left-out* list and *windows* list to store the *left-out* data and the *selected* windows. Let $\sigma_i$ be the computed MI value of $w_i$.

Initially, the search starts from leftmost data, and computes MI value for the first window $w_1 = [s_1, e_1]$. Assuming that $\sigma_1 < \sigma$, then $w_1$ does not satisfy the defined correlation threshold and its indices $[s_1, e_1]$ are inserted into *left-out* list. Next, the search shifts to the right, creating the second window $w_2 = [s_2, e_2]$. The *sliding step* from $s_1$ to $s_2$ indicates how far the window is moved whenever the previous window does not pass the threshold test. The MI is computed for data points belonging to $w_2$. Suppose that this time, $\sigma_2 \geq \sigma$. In this case, the indices $[s_2, e_2]$ are inserted into the *windows* list, and at the same time, the current entry in *left-out* list (i.e., $[s_1, e_1]$) is updated to $[s_1, s_2]$, indicating that only the data partition from $s_1$ to $s_2$ is left out. Next, the search moves on to the third window $w_3 = [s_3, e_3]$ where $s_3$ is right after $e_2$ of $w_2$. The procedure repeats like this for the rest of data series.

Fig. 4 illustrates the results after the first scan. A set of non-overlapping windows which hold significant correlations is stored in the *windows* list, and a set of disjoint data partitions containing left out data is stored in the *left-out* list. These left out data partitions will become the inputs for the next scan with finer granularity.

### 4.2.3 Boxed-Assisted Algorithm with Incremental Computation

Up to this point, we have illustrated how the sliding windows with layered top-down approach can be applied to search for multi-scale correlations along data series. In this section, we describe how to compute MI value for each window in an incremental manner.

As discussed, the *KSG* estimator aims to estimate MI based on the neighborhood population. For each data point $i$, it first searches for $k$-nearest neighbors, and then counts the marginal points within $k$-nearest distance [40]. Different methods can be used to search for $k$-nearest neighbors (e.g., k-D tree, boxed-assisted, projection method [42]). We choose to use the boxed-assisted method because it outperforms the others, especially for low dimensional data [42].

Moreover, notice that the shift from $w_1$ to $w_2$ in Fig. 3 results in three different sets of data. *Set 1* from $s_1$ to $s_2$ contains data points of $w_1$ to be removed from $w_2$, *Set 2* from $s_2$ to $e_1$ is overlapping data, and *Set 3* from $e_1$ to $e_2$ is newly added data. To minimize the computation cost, we design an optimized boxed-assisted algorithm to track changes introduced by removing old data ($[s_1, s_2]$) and by adding new data ($[e_1, e_2]$). The optimization ensures that for each new window $w_2$, only new data need to be computed, and only data affected by the changes are re-evaluated.

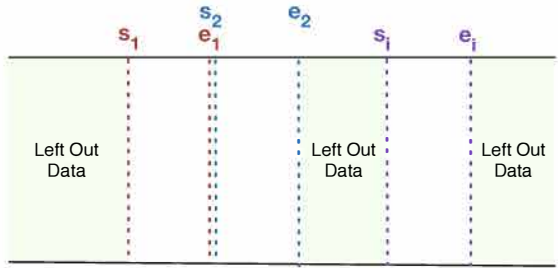

Fig. 4: Time series after the first scan

In the standard boxed-assisted algorithm, the search space is divided into equal size boxes. Each data point is projected into exactly one box. Each box maintains a list storing points belonging to that box. When searching for the $k$-nearest neighbors of point $i$, first the box containing point $i$ is found, then the search starts from that reference box and extends to its neighborhood until the $k$ nearest points are found. Next, the distances (in each dimension) to the $k^{th}$-neighbor are determined and the marginal points are computed by counting the number of points fallen within these distances. In addition to the boxed-array used in the standard version, we use additional data structures to keep track of the previous computation (e.g., of $w_1$). For each data point $i$, we add 3 components to its data structure to store: (1) the index of the $k^{th}$-nearest neighbor, (2) the distance in each dimension to its $k^{th}$-nearest neighbor, (3) the number of marginal points in each dimension. To track the changes caused by old and new data, we introduce the concepts of *influenced region* and *influenced marginal region* for each data point.

*Definition 6.* An *influenced region* (IR) of point $i$ is a rectangular bounding box $R_i = (l_i, r_i, b_i, t_i)$, where $l_i, r_i, b_i, t_i$ are its left-, right-, bottom-, and top-most indices, respectively.

*Definition 7.* The *influenced marginal region(s)* (IMR) of point $i$ is the marginal region(s) located within the nearest distance $d_i$ in each dimension.

To determine the *influenced region*, first the location of point $i$ in its corresponding box is located, $p_i = (x_i, y_i)$. Then the bounding box is formed by expanding from $p_i$ to the distance $d_i = (d_x, d_y)$ of the $k^{th}$-neighbor, i.e. $(x_i \pm d_x, y_i \pm d_y)$. The *IR* maintains an area where any point $j$ either falling into or being removed from this region will affect point $i$. The potential changes in this region can be either altering its $k^{th}$-neighbor or changing its marginal counts. In this case, point $i$ requires a re-evaluation. Instead, the *IMR* maintains an area where any point $j$ either falling into or being removed from it will change the marginal counts of point $i$. When evaluating the newly added and oldly removed points, the *IR* helps to determine when an existing point has to be re-evaluated (both the $k$-nearest neighbor and the marginal counts), and the *IMR* helps to determine when an existing point has to recount its marginalized neighbors.

Fig. 5 illustrates the *influenced region* and *influenced marginal region* concepts, and explains how they can help to minimize computational cost. Consider a data set of seven data points $p_0, ..., p_6$ with their locations projected into boxed-array as in Fig. 5a. Let $p_0$ (in red) be the reference point under monitoring, $k = 2$ be the nearest neighbor parameter, and the *maximum norm*[1] be the distance metric between neighbors. Under this setting, the two nearest neighbors of $p_0$ are $p_1$ and $p_2$ (in green), and its nearest distances in each dimension are $dx$ and $dy$. The nearest distances allow the algorithm to form the marginal regions, from which the marginal counts are computed. In this case for point $p_0$, the marginal counts are $n_x = 3$ (including $p_1, p_2, p_4$), and $n_y = 3$ (including $p_1, p_2, p_3$), respectively. The *influenced region* of $p_0$ is the rectangular colored in green, and the *influenced marginal regions* are those with gray shade in either dimension.

Fig. 5b illustrates how changes are introduced and managed. For simplicity, we only discuss cases when new points are added into the previous computation. Changes introduced by removing points can be handled in similar ways. Assuming that at time $t_1$, a new point $p_7$ is added to the current window. The addition of $p_7$ can create two different types of changes:

– It changes the $k^{th}$-nearest neighbor of $p_0$ (as in Fig. 5b). In this case, a new nearest neighbor search for $p_0$ is required.

– It does not alter the $k^{th}$-nearest neighbor, but increases the marginal count(s) of $p_0$. In this case, no new search is required but only a re-evaluation of marginal count(s).

At time $t_2$, a new point $p_8$ arrives and falls into the $y$-marginal influenced region of $p_0$, for which it will alter the marginal count $n_y$ (but no new search is required in this case). Similarly, a new point $p_9$ will increase the marginal count $n_x$.

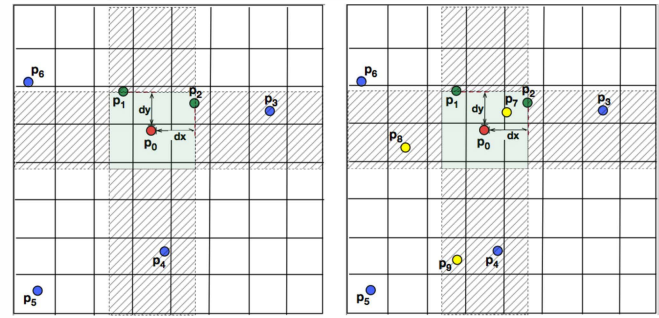

(a) Illustration of IR (green) and IMR (gray shade)

(b) Manage changes in IR (green) and IMR (gray shade)

Fig. 5: Box-assisted algorithm with incremental computation

With the introduction of *IR* and *IMR*, the MI computation is enhanced as follows. For each data point $i$, if $i$ is:

- A new point: (Step 1.1) Follow the standard algorithm to compute its marginal points (Algorithm 1, lines $8 - 10$). (Step 1.2) Re-evaluate every point $j$ whose influenced regions contain $i$ (Algorithm 1, lines $14 - 22$).
- A removed point: (Step 2.1) Remove point $i$ and its corresponding data structures (Algorithm 1, lines $11 - 13$). (Step 2.2) Re-evaluate every point $j$ whose

---

1. $L_\infty$: $d(p_i, p_j) = \| p_i, p_j \|_{max} = max(\| x_i - x_j \|, \| y_i - y_j \|)$

- influenced regions contain $i$ (Algorithm 1, lines $14 - 22$).
- In the overlapping region of two windows, no computation is required.

As the result of this *incremental computation* method, for each window, only a minimum search region (containing new points) and a minimum update region (containing points affected by added and removed points) require additional computation. Algorithm 1 provides the outline of this incre-mental layered top-down search procedure.

---

**Algorithm 1** Layered Top-Down Search with Adaptive Sliding Windows and Incremental Computation
---
    **function** LayeredTopDownSearch $(\{X, Y\})$
    **Input:** $\{X, Y\}$: pair of time series variables
    **Params:** $k$: nearest neighbors parameter,
             $\sigma$: correlation significance threshold,
             $g_{max}$: maximum granularity,
             $g_{min}$: minimum granularity,
             *slide*: shifting step between windows
    **Output:** *windowsList*: list of *selected* windows
1:  *initialize:* $g_c \leftarrow g_{max}$          ▷ $g_c$: current granularity
2:     $leftOutList \leftarrow \{X, Y\}$     ▷ list of *left out* data partitions
3:  **while** $g_c \geq g_{min}$ **do**
4:     **while** notEmpty(*leftOutList*) **do**
5:         $leftOut \leftarrow leftOutList.next$  ▷ gets the next *left out* partition
6:         $window \leftarrow$ getNextWindow($leftOut, g_c$)    ▷ gets the next window in chronological order from the current *left out* partition
7:         **while** $window.endIndex \leq leftOut.endIndex$ **do**
8:             **for** $point_i$ in $window$ **do**
9:                 **if** $point_i$ is a new point **then**
10:                    searchKNeighbors($point_i, k$)
11:                    updateMarginalCounts($point_i$)
12:                **else if** $point_i$ is an old point **then**
13:                    removePoint($point_i$)    ▷ remove $point_i$ from its corresponding data structures
14:                **end if**
15:                $affectedPoints \leftarrow$ getAffectedPoints($point_i$)  ▷ gets all points affected by $point_i$
16:                **for** $point_j$ in $affectedPoints$ **do**
17:                    **if** k-neighbor changes **then**
18:                        searchKNeighbors($point_j$)
19:                        updateMarginalCounts($point_j$)
20:                  **else if** marginal counts change **then**
21:                      updateMarginalCounts($point_j$)
22:                **end if**
23:              **end for**
24:           **end for**
25:           $mi \leftarrow$ computeMI($window$)    ▷ compute MI magnitude
26:           **if** $mi \geq \sigma$ **then**
27:                $windowsList$.insert($window$)
28:                $window \leftarrow$ getNextWindow($leftOut, g_c$)
29:           **else**
30:                $window \leftarrow$ shiftWindow($leftOut, g_c, slide$)
31:                update($LeftOutList$)
32:           **end if**
33:         **end while**
34:     **end while**
35:     $g_c \leftarrow g_c - slide$
36: **end while**
37: **return** $windowsList$
    **end function**

---

*Space complexity:* The space complexity of the standard boxed-assisted algorithm is $O(n)$ where $n$ is the number of samples in a window [43]. With $m$ additional data structures used in the optimized version, the space complexity is $O(mn)$, and thus linear in the data size.

*Time complexity:* With the standard boxed-assisted algorithm, the time to compute MI for a window is $O(n \log(n))$ where $n$ is the window size [43]. The sliding window approach with $w$ windows results in the complexity of $O(wn \log(n))$. Note that the value of $w$ partially depends on the correlation threshold $\sigma$. With the incremental computation, however, it only requires to have computation on new data points and updates on affected points. For example, if the data are sparse, i.e, few overlapping points are affected through insertions of new and removal of old points, the time complexity will be much smaller.

### 4.3 Data Partitioning and Recursive Parallelism using Apache Spark

We exploit the distributed computational capability of the Apache Spark to address the scalability requirement and accelerate the search process. Using the divide-and-conquer strategy, we divide the time series into multiple overlapping data partitions. Each partition is distributed and analyzed on a separate worker node of the Spark cluster. The overlapping data ensures the analysis is contiguous between partitions and is equal to the maximum size of a window. In order to leverage Spark platform for task parallelism, we design a recursive parallel algorithm that work recursively on the partitions. The algorithm is composed of two phases: the *Map* phase and the *Reduce* phase. In the *Map* phase, each data partition is mapped to a worker node where the top down search procedure will be executed to search for significant correlations on that partition. The results returned from this phase include a set of *selected* windows, and a set of *left-out* data. The *Reduce* phase collects these results and does three things: (1) First, it stores the list of *selected* windows to the result collection; (2) Next, it aggregates the *left-out* data from all worker nodes, and recursively invokes the *Map* phase to re-distribute the collected *left-out* data to worker nodes where new search process will start. This recursive procedure stops when either the *left-out* data returned from all worker nodes are empty, or the granularity reaches the *minimum*; (3) And finally, it performs a post processing step, sorting the extracted windows in increasing order of the start index and decreasing order of window length in order to merge overlapping windows and ensure bigger windows will be reported.

The recursive parallel algorithm is implemented in Python using the *PySpark* interface of the Spark platform. The layered top-down search algorithm, however, is implemented in C++ for better performance. Since the search algorithm involves $k$-nearest neighbor search which is often computationally expensive, C++ is a good candidate to perform this search. To communicate between the two programs, we use Spark's *pipe()* command.

Algorithm 2 illustrates this recursive parallel procedure. In lines 1-2, data indices are partitioned into overlapping partitions. Each partition contains $n$ windows and two consecutive partitions are overlapped by a window of granularity $g_{max}$. Lines 4-13 define the *map* and *reduce* phases of the recursive parallel search. The external C++ program performing layered top-down search on each data partition is invoked in lines 14-16.

We prove the correctness of the data partitioning strategy and the parallel algorithm using Lemma 1 as follows.

**Lemma 1.** *Let $(X, Y)$ be a pair of time series of length $N$, and $w$ be any window of granularity $g$. Suppose $(X, Y)$ is divided into*

**Algorithm 2** Recursive Parallel Top-Down Search

```
function RecursiveParallelSearch ({X,Y})
Input: {X,Y}: pair of time series variables
Params: k: nearest neighbors parameter,
        size: window size at granularity g_max,
        n: number of windows in each partition,
        N: total number of samples
1:  sc ← SparkContext(conf=SparkConf());  ▷ initialize SparkContext
2:  dataRdd ← sc.parallelize([(i-size,i+n*size) for i
             in xrange(0,N,n*size)]);  ▷ divide time series into
    overlapping partitions.
3:
4:  def Mapper(dataRdd):
5:    (windows,leftOut) ← dataRdd.map(computeMI).cache();  ▷
      map each data partition to a worker node, and invoke computeMI
      function to search for window-based correlation
6:
7:  def Reducer(windows, leftOut):
8:    selectedWindows ← selectedWindows.union(windows)  ▷
      union selected windows from all worker nodes
9:    selectedWindows.saveAsTextFile('hdfs://')  ▷ store the
      selected windows into the result collection
10:   leftOutData ← leftOut.aggregate(([],0),
            (lambda acc, value: (acc[0].union(value), acc[1] + 1),
            (lambda acc1, acc2: (acc1[0].union(acc2[0]), acc1[1] +
            acc2[1]))))  ▷ aggregate leftout data from all worker nodes
11:   if(leftOutData[1] > 0 AND g_c ≥ g_min):
12:     Mapper(leftOutData)  ▷ recursively invoke the map phase
        to re-distribute leftout data
13:   postProcessing(selectedWindows)  ▷ sorting and merging
      overlapping windows

14: def computeMI(partitionRdd):
15:   (windows,leftOut) ← partitionRdd.pipe(topDownSearch)  ▷
      invoke top-down search procedure in C++ on each data partition
16:   return (windows,leftOut)
```

$p$ partitions which are pairwise overlapped by a window of granularity $g_{max}$. Then the window $w$ is reported iff $I(X_w, Y_w) \geq \sigma \land \nexists w'$ of granularity $g' : w \subset w' \land I(X_{w'}, Y_{w'}) \geq \sigma \land g_{min} \leq g \leq g' \leq g_{max}$.

*Proof.* We will prove that Lemma 1 holds for Algorithm 1 that runs sequentially on a single node, and for Algorithm 2 that runs on multiple overlapping data partitions in a distributed environment, and thus, the correctness is preserved by both algorithms.
◦ If $p = 1$: $(X,Y)$ has only one partition. On $(X,Y)$, Algorithm 1 performs a breadth-first search on all intervals with the defined granularities, starting from the coarsest granularity $g_{max}$. Thus, $w$ will be reported for the coarsest (topmost) granularity where $I(X_w, Y_w) \geq \sigma$ holds. Because $(X, Y)$ has only one partition, Algorithm 2 launches and applies Algorithm 1 to the entire partition, and thus, reports the same results.
◦ If $p = 2$: $(X, Y)$ is divided into two partitions $(p_1, p_2)$ overlapped by a window $w_o$ of granularity $g_{max}$. Let $w_1, w_2$ be any windows in partitions $p_1$ and $p_2$, respectively, that fall into the overlapping region $w_o$. Then:
  – Case a. If $w_1 = w_2 = w_o$:
    * If $I(X_{w_o}, Y_{w_o}) \geq \sigma$, then $w_1$ and $w_2$ are both reported respectively for $p_1$ and $p_2$. The post processing step at the end of the search will merge the two windows into one. On partition $p_2$, the search moves on with the next window right after $w_2$, and continues on the rest of $p_2$ as on the whole of $(X, Y)$.
    * If $I(X_{w_o}, Y_{w_o}) < \sigma$, then neither $w_1$ nor $w_2$ is reported. The search moves on with partition $p_2$ where the next window is shifted from $w_2$ by a shifting step *slide*.
    In both cases, nothing is lost at the overlapping region. Thus, the search is contiguous on $p$ partitions as on $(X, Y)$.
  – Case b. If $w_1 \subset w_2 \subseteq w_o$:
    * If $I(X_{w_1}, Y_{w_1}) \geq \sigma$, then $\exists w_2' \in p_2 : w_1 \subseteq w_2' \subseteq w_2 \land I(X_{w_2'}, Y_{w_2'}) \geq \sigma$. The windows $w_1, w_2$ are both reported on partitions $p_1$ and $p_2$, respectively. The post processing step will merge the two windows at the final stage of Algorithm 2. The search continues on the remaining data of $p_2$ as on the whole of $(X, Y)$.
    * If $I(X_{w_1}, Y_{w_1}) < \sigma$, then either the search finds nothing at the overlapping region if $\nexists w_2' \in p_2 : w_2' \subseteq w_2 \land I(X_{w_2'}, Y_{w_2'}) \geq \sigma$, or it will report $w_2'$ if $I(X_{w_2'}, Y_{w_2'}) \geq \sigma$, and the search continues with the remaining data of $p_2$ as on the whole of $(X, Y)$.
  – Case c. If $w_2 \subset w_1 \subseteq w_o$: this case will not happen because on $p_2$, the search will always start with a window of $g_{max}$. Thus, it will always be either Case a or Case b.
◦ If $p > 2$: every two consecutive partitions in $p$ will be handled similarly to the case of $p = 2$. Thus, $w$ will be always reported if $I_w$ is above the threshold, and no windows are lost at the overlapping region. □

In summary, Lemma 1 shows that the data partitioning strategy guarantees the results are preserved in the overlapping region, and the search is contiguous on $p$ partitions as on the whole of $(X, Y)$.

### 4.4 Setting the Correlation Threshold

When searching for correlations in sliding windows, it is important to have criteria that can justify the strength of an association, as a weak association might not be in the interest of users. The magnitude of mutual information represents how strong the correlation is (the larger the MI, the stronger the correlation), and thus can be used to measure correlation strength. As MI magnitude represents how much uncertainty is reduced between variables, expressing in relation with the entropies, we get:

$$I(X, Y) = H(X) + H(Y) - H(X, Y) \quad (3)$$

From Eq. 3, $I(X, Y)$ is the difference between the sum of the marginal entropies (representing the uncertainty/randomness of the individual variables) $H(X)$ and $H(Y)$, and their joint entropy $H(X, Y)$. Since the joint entropy $H(X, Y)$ is at least as large as each marginal entropy, i.e., $H(X, Y) \geq \max(H(X), H(Y))$, the MI $I(X, Y)$ is bounded by:

$$0 \leq I(X, Y) \leq \min(H(X), H(Y)) \quad (4)$$

Eq. 4 shows that while MI's lower bound is 0, its upper bound varies with the marginal entropies, and can grow unbounded depending on the data size and the underlying data distribution. Thus when data characteristics and their relationships are unknown, it is challenging to set an appropriate threshold to uncover interesting correlations. Having methods to assist users in setting the threshold is necessary, and more importantly, these methods have to provide a quantitative basis that can enable comparison between different correlations. In the following sections, we propose three different approaches to define the threshold $\sigma$, and discuss the advantages and disadvantages regarding their usage.

### 4.4.1 Absolute MI Magnitude using Data Coverage

In this *naive* approach, the threshold $\sigma$ is defined using an absolute real value of MI. The higher this value, the stronger the relationship.

*Definition 8.* A window $w_{X,Y}$ is said to contain significant correlation if its MI value ($\sigma_w$) is greater than or equal to a predefined real number $\sigma$ representing an MI magnitude.

The value of $\sigma$ decides how many windows will show up in the results. Clearly, a larger value of $\sigma$ results in fewer windows. We propose to use *data coverage* as a basis to determine the threshold $\sigma$:

$$DataCoverage = \frac{\#samples\_in\_selected\_windows}{\#total\_samples} \quad (5)$$

*Data coverage* represents the amount of data covered in selected windows, and can be interpreted as the percentage of interesting data over the entire series. For example, consider the pair of time series $\{X, Y\}$ with size $N = 500$, and assume that the selected windows returned by AMIC contain 100 samples in total. This results in the coverage: $100/500 = 0.2$, accounting for 20% of data. A coverage of 20% implies that significant correlations are expected to reside within 20% of the entire data series. Using *data coverage*, users will tune the $\sigma$ value so that returned windows provide a desired coverage. Since this approach uses a real value of MI magnitude to assess correlation in a window, it can reveal the real strength of a relationship. However, it is a trial-and-error approach because *data coverage* will require a tuning step in order to find an appropriate value for $\sigma$. Thus, in practice this approach is difficult to use if no prior knowledge of the considered data is available.

### 4.4.2 Two-Step Filtering using Normalized Entropy and Normalized Mutual Information

Alternatively, we propose a two-step filtering approach to select a window based on *normalized entropy* and *normalized mutual information*. Normalization scales unbounded real values to a range of relative values, in this case is $[0, 1]$, and thus has the advantage of providing comparative and bounded values when setting a correlation threshold.

Consider a window $w_{X,Y} = \{(x_1, y_1), ..., (x_n, y_n)\}$ of $n$ samples obtained from a pair of time series variables $(X, Y)$. The entropy of window $w$ is:

$$H_w = -\sum_{y \in Y_w} \sum_{x \in X_w} p(x,y) log(x,y) \quad (6)$$
$$= H(X_w) + H(Y_w) - I(X_w, Y_w) \quad (7)$$

Window $w$ can obtain a maximum entropy when $X$ and $Y$ are uniformly distributed on $[X_w \times Y_w]$, as uniform distribution provides the largest uncertainty (or equivalently, the least information) over discrete variables [44]. The maximum entropy of window $w$ is:

$$\max(H_w) = \log(n) \quad (8)$$

where $n$ is the window size. In turn, its normalized entropy is:

$$0 \leq \tilde{H}_w = \frac{H_w}{\max(H_w)} = \frac{H_w}{\log(n)} \leq 1 \quad (9)$$

The window entropy refers to the amount of uncertainty contained in the window, thus when its normalized entropy scales to 0, the considered data has no degree of randomness (i.e., is deterministic). In contrast, when $\tilde{H}_w = 1$, the window attains the maximum degree of uncertainty (i.e., data are uniformly distributed). In order to report any meaningful correlations, data in the window have to contain enough randomness ($\tilde{H}_w > 0$) to avoid the deterministic situation. Using the concepts of max entropy and normalized entropy, the window's MI can be normalized in two distinct ways.

*A. Mutual Information Normalized by Max Entropy:* In this approach, maximum entropy of the window is used to normalize its mutual information:

$$0 \leq \tilde{I}_w^{(1)} = \frac{I_w}{\max(H_w)} = \frac{I_w}{\log(n)} \leq 1 \quad (10)$$

By dividing the window's MI by its maximum uncertainty, $\tilde{I}_w^{(1)}$ represents the minimum proportion of information that can be obtained between two variables of interest. Under this approach, windows that have the same size are comparable, and the one which has the larger $\tilde{I}_w^{(1)}$ is preferable.

*B. Mutual Information Normalized by the Window Entropy:* In this approach, the window entropy is used to normalize its mutual information:

$$0 \leq \tilde{I}_w^{(2)} = \frac{I_w}{H_w} \leq 1 \quad (11)$$

By dividing the window's MI by its own entropy, $\tilde{I}_w^{(2)}$ represents the actual reduction of uncertainty of the window. The larger $\tilde{I}_w^{(2)}$ is, the more information is shared between the window's variables, and thus more preferable.

*C. Two-step filtering procedure for window selection:* Let $\sigma_H$ and $\sigma_I$ be the thresholds of normalized entropy and normalized MI, respectively.

*Definition 9.* A window $w_{X,Y}$ is said to contain significant correlation if its normalized entropy ($\tilde{H}_w$) and normalized MI ($\tilde{I}_w$) are greater than or equal to the defined thresholds $\sigma_H$ and $\sigma_I$, respectively.

Finally, the two-step filtering procedure for selecting a window is proposed as follows. For each window $w_{X,Y}$:

- *Step 1*: Select window $w_{X,Y}$ if $\tilde{H}_w \geq \sigma_H$
- *Step 2*: Select window $w_{X,Y}$ returned from *Step 1* if $\tilde{I}_w^{(1)} \geq \sigma_I$ or $\tilde{I}_w^{(2)} \geq \sigma_I$

*D. Discussions* Since $\tilde{I}_w^{(1)}$ and $\tilde{I}_w^{(2)}$ represent two different amounts of shared information, their usage will differ depending on the considered contexts. While $\tilde{I}_w^{(1)}$ allows windows of the same size to be comparable, it cannot compare windows of different sizes. Indeed, $\tilde{I}_w^{(1)}$ gives more preference on small windows over big windows. Thus, the use of $\tilde{I}_w^{(1)}$ is appropriate in contexts where users prefer small size windows over big ones. Note that small windows often represent intensive correlated periods. On the other hand, $\tilde{I}_w^{(2)}$ represents actual shared information, allowing users to compare windows of different sizes. It is suitable for scenarios where windows are preferred to be ranked based on their shared information. We illustrate the usage of $\tilde{I}_w^{(1)}$ and $\tilde{I}_w^{(2)}$ in the following example.

Consider Fig. 6, plotting the time series between the *Wind Speed* and the *Taxi Trips* during a storm event. The event started at the beginning of $28^{th}$ Mar, and ended on $31^{st}$ Mar 2011. Let $w_1$ and $w_2$ be two extracted windows. Since $\tilde{I}_w^{(1)}$ places preference on small windows, it is likely to rank $w_1$ as more *interesting* than $w_2$ ($\tilde{I}_{w_1}^{(1)} > \tilde{I}_{w_2}^{(1)}$). Notice that in this case, $w_1$ represents a more intensively correlated period between the two variables. In contrast, $\tilde{I}_w^{(2)}$ is likely to rank $w_2$ as more *interesting* than $w_1$ ($\tilde{I}_{w_1}^{(2)} < \tilde{I}_{w_2}^{(2)}$), thus capturing the entire storm event from its start to its end.

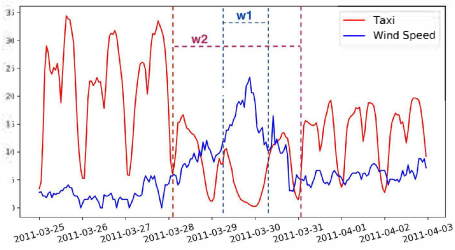

Fig. 6: Illustrate $\tilde{I}_w^{(1)}$ and $\tilde{I}_w^{(2)}$ usage

## 4.5 Positive and Negative Correlation

The AMIC framework relies on MI to identify temporal correlations between variables, and the extracted windows are the time periods where correlations exist. However, a positive MI only guarantees the presence of correlation, but does not reveal whether the correlation is *positive* or *negative*. Distinguishing *positive* and *negative* correlation helps understand the nature of the dependency, as a *positive* correlation means one variable enhances another variable, and a *negative* correlation means one variable inhibits the other. For example, a *positive* correlation between the *Taxi Trips* and the *Number of Collisions* indicates that the increasing (or decreasing) of taxi trips is associated to the increasing (or decreasing) of the number of collisions. A *negative* correlation between the *Taxi Trips* and the *Wind Speed* indicates that the decreasing (or increasing) of taxi trips is associated to the increasing (or decreasing) of the wind speed. We define the following method to determine whether the variables in extracted windows are *positively* or *negatively* correlated.

Consider the window $w_{X,Y} = \{p_1,...,p_n\}$ extracted from AMIC. Since $w_{X,Y}$ satisfies the defined threshold, it contains a statistically significant correlation. Let $p_i = (x_i, y_i)$ and $p_{i+1} = (x_{i+1}, y_{i+1})$ be two *consecutive* data points of $w_{X,Y}$ measured at times $t_i$ and $t_{i+1}$. We say that $X$ and $Y$ are *positively associated* during the period $[t_i, t_{i+1}]$ if one of the following holds:

- $x_i < x_{i+1}$ and $y_i < y_{i+1}$
- $x_i > x_{i+1}$ and $y_i > y_{i+1}$

In other words, during the period $[t_i, t_{i+1}]$, the values of $X$ and $Y$ both increase, or decrease, together. Similarly, we say that $X$ and $Y$ are *negatively associated* during the period $[t_i, t_{i+1}]$ if the value of one variable increases and the value of the other variable decreases:

- $x_i < x_{i+1}$ and $y_i > y_{i+1}$
- $x_i > x_{i+1}$ and $y_i < y_{i+1}$

**Definition 10.** *Positive temporal period* ($\#PP_w$) of the window $w_{X,Y}$ is the number of periods $[t_i, t_{i+1}]$ where $X$ and $Y$ are *positively associated*.

**Definition 11.** *Negative temporal period* ($\#NP_w$) of the window $w_{X,Y}$ is the number of periods $[t_i, t_{i+1}]$ where $X$ and $Y$ are *negatively associated*.

**Definition 12.** The degree of *positive/negative association* between $X$ and $Y$ in the window $w_{X,Y}$ is defined as:

$$-1 \leq \mu_w = \frac{\#PP_w - \#NP_w}{|w| - 1} \leq 1 \quad (12)$$

where $|w|$ is the number of data points in $w_{X,Y}$. Using $\mu_w$, we define the nature of correlation between $X$ and $Y$ as follows. Let $w_{X,Y}$ be the window extracted from AMIC.

*Positive correlation:* The correlation between two variables $X$ and $Y$ in the window $w_{X,Y}$ is *positive* if they are *positively associated* in $w_{X,Y}$, i.e., $0 < \mu_w \leq 1$. The *positive* correlation becomes stronger as $\mu_w$ gets closer to 1.

*Negative correlation:* The correlation between two variables $X$ and $Y$ in the window $w_{X,Y}$ is *negative* if they are *negatively associated* in $w_{X,Y}$, i.e., $-1 \leq \mu_w < 0$. The *negative* correlation becomes stronger as $\mu_w$ gets closer to $-1$.

*Neither positive nor negative correlation:* The correlation between two variables $X$ and $Y$ in the window $w_{X,Y}$ is neither *positive* nor *negative* if their $\mu_w = 0$. This type of correlation occurs in non-linear dependencies, for example, when the relation between $X$ and $Y$ follows non-linear functions such as square or cube.

Note that the case in which correlation is neither *positive* nor *negative* is highly sensitive to noise. For example, the presence of even a few noisy samples in the data would alter the value of $\mu_w$ from $\mu_w = 0$ to $\mu_w \neq 0$. In this case, a *confidence level* can be used to assert the significance of the association:

$$c_p = \frac{\|\#PP_w - \#NP_w\|}{\#PP_w} \quad (13)$$

$$c_n = \frac{\|\#PP_w - \#NP_w\|}{\#NP_w} \quad (14)$$

$$c_o = 1 - \|\mu_w\| \quad (15)$$

where $c_p$, $c_n$, and $c_o$ are the *confidence level* of *positive*, *negative*, and neither *positive* nor *negative* correlations, respectively.

## 5 EXPERIMENTAL EVALUATION

We evaluate the scalability and effectiveness of AMIC using both synthetic and real-world data sets. The generated synthetic data sets contain different types of dependencies (e.g., linear vs. non-linear). The two real-world data sets are obtained from the NYC Open Data Portal [2] and from an energy trading company[2] in Denmark. The effectiveness evaluation aims to assess the interesting and/or useful level of extracted windows. The scalability evaluation asserts the performance and robustness of AMIC. In the following discussion, we describe our infrastructure, data, experiments and findings.

### 5.1 Infrastructure

The synthetic and the NYC data sets are evaluated using the infrastructure and data facility at the Center of Urban Science and Progress, New York University [45]. Our infrastructure includes a 1200-core cluster running Cloudera Data Hub 5.4 and Apache Spark 1.6. The cluster consists of 20 high-end nodes, each with 64 cores, 256GB of RAM, and 24TB of storage. With the energy data sets, due to the NDA signed with the company, we perform the analysis on our local server at Aalborg University [46]. The server has 2.7 GHz processor, 8 GB of RAM, and 250 GB of storage.

2. Due to the Non Disclosure Agreement (NDA) with the company, we do not disclose the company name and its data sets.

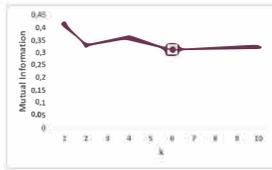

Fig. 7: Mutual Information with different k

## 5.2 Parameters Setting

– *Value of k:* We use $k$ ranging from 1 to 20 to compute MI for the variables extracted from the real data sets. The MI values produced by different $k$ are compared together. We found that $k$ between 1 and 4 produces high variance of MI value, while the MI becomes more stable with $k$ from 5 to 10. The value $k = 6$ gives the most stable result, thus, is selected to be used in AMIC. The tuning process is illustrated in Fig. 7, which shows that starting from $k = 6$, the MI is converged to a similar value.

– *Threshold $\sigma$:* This parameter is closely related to the nature of the data. For those that are naturally correlated such as the synthetic data sets (generated with known relations), the desired relationships can be found even with high $\sigma$. For real-world data sets where noise is present and correlations can be rare, a lower $\sigma$ value is helpful to uncover infrequent correlations. We will discuss the value of this parameter when analyzing each specific set of data.

– *Max granularity:* When applying AMIC to real-world time series, we set the granularity to *year*, covering the entire data series in order to test the overall correlation. To search for time windows, we start with the *month* granularity, then split into *week*, *day*, then *hour*. The reduction of the granularity will also depend on the amount of *left out* data returned from the preceding search.

## 5.3 Synthetic and Real-World Data Sets

*The synthetic data sets:* We generate synthetic data sets containing multiple types of known relations, shown in Fig. 8. They include both linear and non-linear dependencies, as well as monotonic and non-monotonic functions. These are standard relations often used to evaluate correlation methods [38], [47]. Figs. 8a and 8b illustrate two independent variables where we add 10% outliers (following a linear dependency) to the latter. Figs. 8c and 8d demonstrate linear relationships, with 10% outliers (uniformly distributed) added to the latter. Figs. 8e, 8f, 8g, 8h, 8i, and 8j represent the exponential, the quadratic, the diamond, the circle, the sine and the cross functions, respectively. We provide the detail of synthetic data sets generation in Appendix A.

*The NYC Open Data [2]:* This source has more than 1,500 spatio-temporal data sets providing a wide range of information about New York City: from its infrastructures and resources (e.g., water, energy) to its environment and citizens (e.g., employment, education, mobility). Exploring these data sets can unveil invaluable insights about the city and the life of its citizens. For evaluation purpose, we test AMIC on two collections of data: the transportation-related and the weather-related data sets [35].

*The Energy Data Sets:* The data are obtained from an energy trading company in Denmark. The data sets provide information of energy produced by different sources such as wind turbines, solar panels and power plants. The data are sampled in minute intervals, and is recorded at production unit level, i.e., energy aggregated for an entire production unit, such as the entire power plant or wind turbine. Respecting the NDA signed with the company, we do not disclose the data sets, but will discuss correlation findings obtained from them.

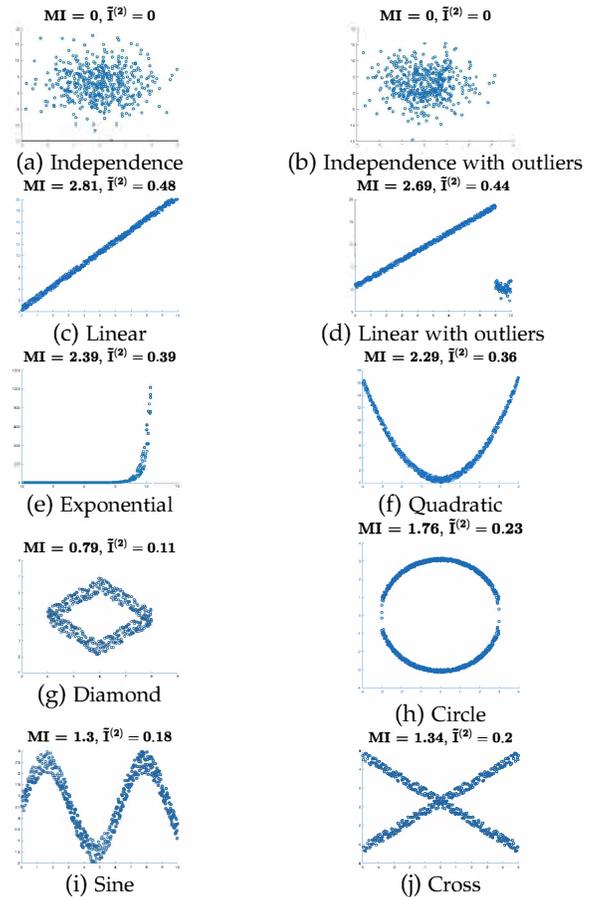

Fig. 8: Evaluation on synthetic data sets

## 5.4 Data Cleaning and Preprocessing

On real-world data sets, we first perform data cleaning by removing duplicated entries and interpolating missing values using linear interpolation [48]. Once the data are cleaned, the preprocessing process is performed to transform raw data into variables of interest, using available meta-data associated with each data set. To deal with large data volumes, the preprocessing process is performed on our Spark cluster. It took approximately 45 minutes to process the taxi data set, and less than 10 minutes to process each of the other data sets.

From the selected NYC Open Data Sets [35] and the energy data sets, we extract the variables of interests considering multiple time resolutions (D: Day, H: Hour, 15M: 15 Minutes, 1M: 1 Minute). These variables will be the inputs of AMIC to test the method's effectiveness on real-world data.

## 5.5 Evaluation of Synthetic Data Sets

The synthetic data sets are used to evaluate the correctness and scalability of the AMIC method, as the known relations provide the ground truth, enabling us to correctly verify whether AMIC can identify the defined relationships. The evaluation is done both separately for each relation and together for multiple relations. Fig. 8 demonstrates the separate evaluation on each relation. The MI magnitude and the normalized MI value $\tilde{I}^{(2)}$ are computed for each of them, shown in the corresponding figures (Fig. 8). As can be seen, their MI magnitudes are significantly larger than 0, indicating that there is dependency among generated data. Specifically, linear relations obtain the highest MI value ($MI > 2$), with a reduction in uncertainty ($\tilde{I}^{(2)}$) up to nearly 50%. The non-linear functional relations (exponential and quadratic) obtain the second largest MI values and uncertainty reduction ($> 30\%$). Finally, the non-linear non-functional relations (diamond, circle, sine, and cross) have the smallest MI values and percentages of uncertainty reduction ($> 10\%$). Notably, comparing the MI magnitudes in Figs. 8c and 8d shows that the added outliers have weakened the linear relationship, and this phenomenon is visible and is detected using AMIC.

To evaluate multiple relations, we combine them into the same data series, following a predefined order. Random noise (i.e., independent data) is added between two different relations to separate them. We apply AMIC to the data in order to search for windows of embodied relationships. Fig. 9 illustrates a combination of four relations: Cross, Diamond, Sine and Quadratic, with data of independent variables added in between. As can be seen, our method is able to separate each relation, returning four distinct windows together with their indices and corresponding MI values. With synthetic data sets, since we already know the data characteristics, we use the absolute MI magnitude with data coverage method (Section 4.4.1) to set the MI threshold.

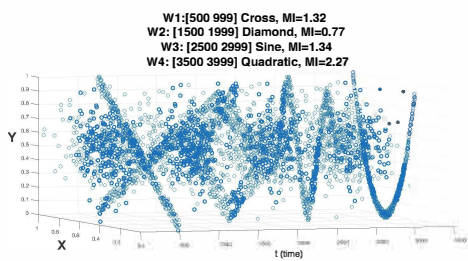

Fig. 9: Evaluation of multiple relations: Cross, Diamond, Sine and Quadratic

## 5.6 Evaluation of Real-World Data Sets

When dealing with real-world data sets, since we do not know the data characteristics in advance, we use the two-step filtering method in Section 4.4.2 to define the correlation threshold. Moreover, since we are interested in comparing and ranking windows of different sizes, we use the window's entropy to normalize its mutual information ($\tilde{I}^{(2)}$). In general, we set $\sigma_H = \sigma_I = 0.2$, selecting window that contains at least 20% of randomness with respect to its max entropy and has mutual information which helps reduce at least 20% of uncertainty.

*A. The NYC Open Data Sets.* We apply AMIC to each pair of extracted variables. The findings are summarized in the following discussion, with only a few extracted windows plotted due to space limitation. When discussing the results, we generally compare them to those reported in [19], which uses the same data sets as ours, and applies a topology-based approach to represent and identify relationships among the data sets.

***Weather and Taxi:*** The findings listed here come from variables extracted from *Weather* and *Taxi* data sets.

TABLE 1: Top ranked windows between Taxi Trips and Wind Speed

| From | To | MI | Event |
|---|---|---|---|
| 2012-Oct-29 | 2012-Nov-02 | 0,651174 | Sandy Hurricane |
| 2012-Jul-27 | 2012-Jul-28 | 0,604444 | Tornado hits NYC |
| 2012-Jan-20 | 2012-Jan-21 | 0,578166 | Snow storm |
| 2013-Jun-07 | 2013-Jun-10 | 0.542931 | Tropical Storm Andrea |
| 2012-Nov-10 | 2012-Nov-11 | 0,542242 | Snow storm |
| 2012-Dec-23 | 2012-Dec-24 | 0,493842 | Storm |
| 2012-Nov-06 | 2012-Nov-07 | 0,487898 | Snow storm |
| 2012-Sep-08 | 2012-Sep-09 | 0,424399 | Tornado |
| 2012-Sep-19 | 2012-Sep-20 | 0,420836 | Storm |
| 2011-Aug-26 | 2011-Aug-31 | 0,4124 | Irene Hurricane |
| 2011-Oct-29 | 2011-Oct-31 | 0,40616 | Snow storm |

***Taxi Trips and Wind Speed:*** We examine the correlation between *taxi trips* and *wind speed* variables, in both one hour and 15 mins resolutions. We found a strong negative relationship between them in both resolutions. When the wind speed readings were exceptionally high, major drops in taxi trips occurred (e.g. Fig. 10). Our finding is similar to the result reported in [19], indicating a negative relationship between the taxi trips and the average wind speed.

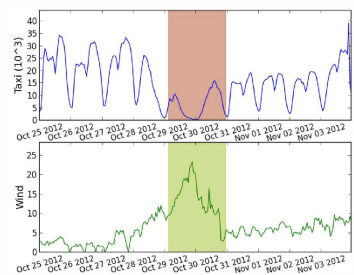

Fig. 10: Taxi Trips vs. Wind Speed

Moreover, as we observe that the correlation often occurred at times when the wind speed had abnormal values, we hypothesize that the discovered relationship might only appear in extreme events. To test this hypothesis, we asked AMIC to rank all extracted windows by their MI magnitudes, and select the top-k windows. As expected, many of these top windows are associated with extreme weather events that had happened in NYC. Table 1 reports some of these events, for which we found their information through archived news feeds. This demonstrates a strength of AMIC which is not mirrored by the current state of the art. Specifically, the ability of AMIC to automatically rank extracted time windows based on their correlation significance has strengthened the data exploration process, and provided users the ability to not only compare different correlations in terms of their strength and significance, but also to determine the exact time periods where the correlations happen. Thus, users are able to make further steps to understand the nature of discovered correlations.

***Taxi Trips and Rain Precipitation:*** We examine the relationship between the *number of taxi trips* and *average rain precipitation*. We found that the two variables are not correlated overall, but only in certain time windows. In the extracted windows, we found a negative relation between these two variables. An example is given in Fig. 11. The two highlighted windows show a drop in the taxi trips associated with abnormally high rain. To better explain the extracted windows, we examine their data. The first window lasts for two days, from 29$^{th}$ Oct 2012 to 30$^{th}$ Oct 2012, the period when hurricane Sandy approached NYC. In the second window, however, the drop of the taxi trips happened at midnight. Note that taxi trips have a daily pattern, showing a high number of trips during rush hours (7AM-9AM, 1PM-3PM, 6PM-8PM) and a low number of trips in non-rush hours, especially at midnight. Thus, the low taxi trips in second window might also be caused by the lack of taxi demand at midnight. In this case, the correlation in second window can be coincidental.

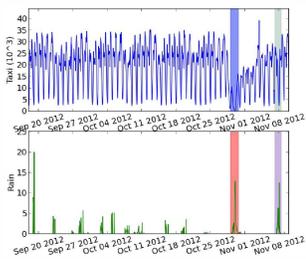

Fig. 11: Taxi Trips vs. Rain Precipitation

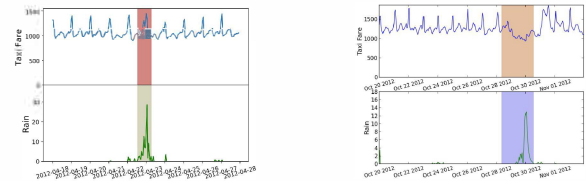

(a) During normal days    (b) During hurricane

Fig. 12: Taxi Fare vs. Rain Precipitation

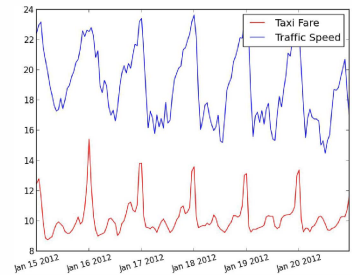

Fig. 13: Daily Patterns of Taxi Fare and Traffic Speed

Moreover, it should also be noted that there is a short period on the far left of Fig. 11, where an abnormally high rainfall is also linked with a drop in taxi trips. This window, however, is not shown in our search results. We found that the abnormal increase in precipitation occurred in a very short time period. Only 7 data samples were reported in this period, thus the window's entropy is lower than the threshold $\sigma_H$, and is not selected by AMIC (Section 4.4.2).

The findings in [19] report a strong negative relationship between the two variables considered here: When rain precipitation is high, the number of taxi trips is low, implying the difficulty to find a taxi in rainy days. Using AMIC, we extend further this result. We found that not only are the two variables negatively correlated in certain time windows, but also the correlations mainly occur during extreme weather events, such as during a hurricane. This allows users to make further investigation, for example, to combine multiple factors such as wind speed and rainfall, and come up with a better explanation for the cause of taxi trips dropping.

*Taxi Fare and Rain Precipitation:* A finding in [19] reports a positive relationship between *Taxi Fare* and *Rain Precipitation*, suggesting that taxi drivers increase earnings when it rains. In our findings, we also confirmed this positive correlation, as an example in Fig. 12a where the highlighted window shows a positive association between high rainfall and a slight increase of taxi fare. More importantly, we found that this positive correlation is visible only in *hour* resolution, but disappears in *day* resolution. This phenomena can be explained due to the fact that taxi drivers are target earners: taxi drivers have a daily income target, and reach their targets sooner when it rains, after which they quit driving for the remaining of the day, thus, maintain their overall income under *day* resolution. Additionally, AMIC shows that the positive correlation between taxi fare and precipitation only occurs during *normal* rainfall, whereas during extreme events such as during a hurricane, a negative correlation is observed instead. This counterexample is shown in Fig. 12b, where taxi fare and precipitation are negatively correlated during the hurricane Sandy. This may be because of a shutdown in the city due to the hurricane.

*Taxi Trips and Visibility:* We found a positive correlation between the *number of taxi trips* and *visibility*. During periods when the visibility is low, a decrease in taxi trips occurs. This may be due to the increased danger of driving under such a condition. In addition, our findings also show that this phenomenon happens mostly at midnight, which suggests that beside visibility, other factors such as the decrease in taxi demand, might also play a role in causing a drop in taxi trips.

**Taxi and Traffic Speed:** We analyze variables extracted from the *Taxi* and *Traffic Speed* data sets. When applying AMIC to the yearly data of the pair (*Taxi Trips*, *Traffic Speed*), we found a strong negative correlation with a significantly high MI magnitude. We then split the data into monthly, weekly and daily patterns, and apply AMIC to them. We also obtain high MI values. This suggests that taxi trips and traffic speed are overall correlated.

We examine another relationship, *Taxi Fare* and *Traffic Speed*, in *hour* resolution. We obtain a relatively high MI for the entire data series, implying an overall correlation between taxi fare and traffic speed. Particularly, when we split the yearly data into a daily pattern, we gain even higher MI values, implying the correlation is stronger in daily pattern. Fig. 13 shows this daily pattern, where the traffic speed and the taxi fare were both low and high at similar times. This suggests that taxi drivers are likely to earn less with slow traffic.

We analyze the pair (*Taxi Fare*, *Trip Duration*), and found a slightly weak positive correlation between them. The correlation suggests that the taxi driver is likely to earn more when the trip duration is longer. Note that trip duration gets longer either because of traffic jam (the fare does not increase) or because the travel distance is long (the fare does increase). If traffic jam is the cause of long trip duration, linking *trip duration* to *taxi fare* is not enough, and thus explains the *weak* correlation between the two variables. This might suggest that other variables, such as *travel distance* or *traffic speed*, can be of interest in analyzing together with *taxi fare*. Moreover, we also found that the correlation between

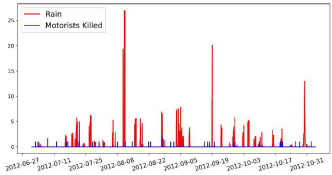

Fig. 14: Rainfall vs. Motorists Killed

*taxi fare* and *trip duration* in *day* resolution is stronger than in *hour* resolution. This can be because long trip duration often lasts longer than an hour, thus the association between high *taxi fare* and long *trip duration* (and vice versa) is more visible under *day* resolution.

**Collisions and Weather:** We analyze the *Collisions* and *Weather* data sets. One of the findings in [19] reports a strong positive relationship between the rainfall and the number of motorists killed. However, AMIC does not confirm this relationship. The MI value between this pair of variables is almost zero, and no significant windows are extracted from the data series, even with a low threshold $\sigma_I$, across all resolutions. Fig. 14 shows a part of the time series between the two variables (in *day* resolution), where no associated patterns can be seen in the figure: high (or low) rainfall seems to have no effects on the number of motorists killed.

Similarly, AMIC does not confirm the strong positive relationship between the *rainfall* and the *number of injured pedestrians*, as reported in [19]. We instead observe random patterns between the time series of these two variables.

**Collisions and Taxi:** We study variables extracted from the *Collisions* and the *Taxi* data sets. In [19], a strong positive relationship between the *Taxi Trips* and the *Number of Collisions* is suggested in its findings. However, our findings are slightly different. We found that overall, the two variables are not correlated, except certain periods during the week displayed a weak positive association. Fig. 15 shows an example of the extracted windows. The highlighted window shows a time period where a high number of taxi trips is associated with a high number of collisions. Intuitively, one can argue that there can be a causal relation between taxi trips and collisions: a high number of taxi trips leads to a high number of collisions. This pattern is visible in the extracted windows from AMIC. However, there are also counter examples (can be seen in Fig. 15), where a high number of taxi trips are also associated with a fluctuation in the number of collisions. This result could suggest that, although there is a positive correlation between *taxi trips* and *number of collisions* in the extracted windows, there might be other hidden variables that can be the main causes of the increase in collisions and that happens during periods when the number of taxi trips is high. This result opens new opportunities for users to look deeper and further investigate their data sets.

**Collisions and 311 Complaints:** We examine the relationships between the *Collisions* and the *311 Complaints* data sets. The *311 Complaints* data set records complaints made by NYC citizens through the 311 phone number.

A finding in [19] suggests that there is a strong positive relationship between the *Number of Collisions* and the *Number of 311 Complaints*. However, AMIC does not confirm this strong correlation. Similar to the case of *Taxi* and *Collisions* data sets, AMIC found that the numbers of collisions and 311 complaints are not correlated overall, but a weak positive correlation is found in the extracted windows. This might suggest the presence of hidden variables in the periods where these two are weakly correlated. Additionally, our findings suggest that the number of complaints made by 311 calls has a daily periodic pattern, where the complaints are significantly higher at late night, and low at other time periods. This could be a consequence of the NewYorkers' lifestyle pattern, for example, noise tends to increase during late night.

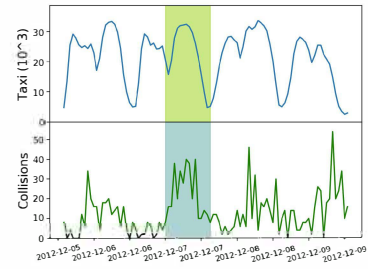

Fig. 15: Taxi Trips vs. Collisions

**B. The Energy Data Sets.** We analyze the extracted variables from the energy data sets. We found a strong positive correlation between the amount of produced energy and the source that generates it. For example, when solar panel is the energy source, a strong positive correlation between *Energy Production* and *Solar Irradiation* is found, in both horizontal and tilted directions (i.e., the *Horizontal Irradiance* and the *Tilted Irradiance*). Whenever the solar irradiation is high, the generated energy increases, and vice versa. As solar panels rely on solar cycle to generate energy, this strong correlation holds for the entire data series of the considered variables.

Similar findings are also obtained when wind turbine is the energy source. A strong positive dependency between *Wind Speed* (as well as *Wind Speed EastWest* and *Wind Speed NorthSouth*) and *Energy Production* is suggested by AMIC. Additionally, we found that *Wind Speed*, *Wind Speed EastWest* and *Wind Speed NorthSouth* all have a relatively similar degree of correlation strength with *Energy Production*. We obtained $\tilde{I}^{(2)} \approx 0.3$, i.e., approximately 30% uncertainty is reduced between *Energy Production* and the other three variables.

When solar is the energy source, we found a negative correlation between *Energy Production* and *Humidity*, and a positive correlation between *Energy Production* and *Temperature*. Whenever the humidity is low, the produced energy is high and vice versa. In contrast, the amount of produced energy increases as the temperature increases. This leads us to examine the relationship between *Humidity* and *Temperature*. As expected, a negative correlation between these two variables is confirmed. The associations between the three variables are naturally intuitive, since high solar irradiance and low humidity result in high temperature, and vice versa.

When wind turbine is the energy source, we found a weak negative correlation between *Air Pressure* and *Energy Production*, as well as a weak negative correlation between *Air Pressure* and *Wind Speed*. These correlations can be intuitively explained. As the air flows from high pressure area to low pressure area, and thus creates wind, it results in a negative correlation between air pressure and wind speed.

Since the wind speed is directly correlated to the produced energy (through the wind turbine), it explains the negative association between air pressure and the energy production.

Similarly, when examining other weather-related variables, we found a positive correlation between *Wind Speed* and *Temperature*, and a negative correlation between *Wind Speed* and *Humidity*. These two correlations are also intuitive. The difference in temperature between two geographical locations creates the motion of particles, which in turn creates wind, thus the higher the difference in temperature, the stronger the wind speed. This explains a strong co-dependence between these two variables. Similarly, stronger winds lead to an increase of evaporation, thus more humidity, and therefore results in a negative correlation between them. Note, however, that while we discuss the correlations between weather-related variables found by AMIC, we do not imply the order of causality between them. Determining causality in the relations between variables is outside the scope of this work.

We test another pair of variables, *Wind Direction* and *Wind Speed*. Interestingly, we found a relatively weak positive correlation between them. As *Wind Direction* indicates the direction in which the wind flows (measured in degrees), and *Wind Speed* describes how fast the air is moving (measured in km/hour), it is not obvious how to explain the nature of this correlation. However, the discovered correlation might prompt users new opportunities for further analysis. As we do not attempt to explain the causality behind the correlation, we refer readers to other articles such as [49] for further understanding of this phenomenon.

Analyzing the data sets further, we examine the energy estimation models used by the company. Since the energy trading company has to manage and trade the produced energy, it needs to estimate the energy production based on its resources (such as solar and wind). To understand how accurate the estimation model is, we analyze the pair (*Energy Production*, *Estimated Production*). As returned by AMIC, the two variables are strongly correlated, with the MI magnitude between the pair is nearly equal to their joint entropy, indicating that the two variables are almost identical. This verifies that the energy estimation model used by the company provides highly accurate estimated values.

Beside discovering correlations, AMIC also prunes multiple pairs of variables that are not correlated. For example, AMIC shows that there is no correlation between (*Energy Production*, *Estimated Running Capacity*), (*Energy Production*, *Nacelle Direction*), or (*Estimated Reduction*, *TSO (OnShore)*). These pruning results show the reliability of AMIC, as one should not expect any dependencies between the produced energy and the engine direction.

### 5.7 Threshold Effects

To evaluate the effects of the correlation threshold, we combine synthetic relations (generated in Section 5.3) into a single data set and apply AMIC using different thresholds. In Figure 16, *Total Windows* are the total number of extracted windows satisfying the threshold, and *Meaningful Windows* are the extracted windows that contain the defined relations. Figure 16 shows that when the threshold is too low (e.g., $\leq 0.3$), it can produce many uninteresting windows, and when the threshold is too high (e.g., $\geq 0.6$), it might miss interesting correlations. These results use $\tilde{I}_w^{(2)}$ to set the threshold, although using $\tilde{I}_w^{(1)}$ produces similar results.

TABLE 2: Comparison between different correlation metrics

| Relations | AMIC (MI) | PCC | dCor |
|---|---|---|---|
| Linear | ✓ (2.689) | ✓ (0.998) | ✓ (0.998) |
| Exponential | ✓ (2.391) | ✓ (0.461) | ✓ (0.584) |
| Quadratic | ✓ (2.291) | ✗ (0.01) | ✓ (0.49) |
| Diamond | ✓ (0.793) | ✗ (-0.03) | ✓ (0.228) |
| Circle | ✓ (1.755) | ✗ (0) | ✓ (0.118) |
| Sine | ✓ (1.295) | ✗ (-0.07) | ✓ (0.377) |
| Cross | ✓ (1.34) | ✗ (0) | ✓ (0.245) |

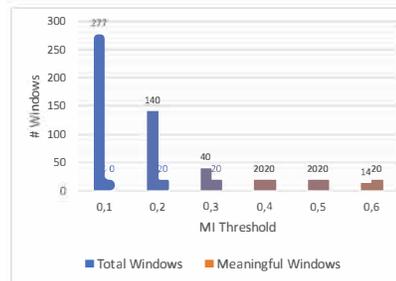

Fig. 16: Threshold Effects

### 5.8 Comparison Against Other Methods

Comparison between different correlation metrics has been studied in the literature, such as in [38]. In this section, we make a short comparison between our method (i.e., AMIC using MI) and those using two other representative metrics: the Pearson correlation coefficient (PCC) [36] and the distance correlation *dCor* [50]. We apply these metrics on synthetic data sets to verify their capability in discovering different types of relations. The results are reported in Table 2 (a check mark indicates the correct identification of the corresponding relation, while a cross indicates non-identification, and the number in each column is the value computed by the corresponding metric). In this table, PCC can only identify two relations (with the check mark), while MI and *dCor* can identify all of the relations. Although *dCor* has similar capability as MI in discovering different types of relations, it can work only with numeric values (because of the distance measure). In contrast, MI can work with any types of data because it relies on the probability distribution, and thus fits better in Big Data contexts and is a better candidate to address the variety challenge of Big Data.

### 5.9 Performance and Scalability Evaluation

*Performance evaluation:* To evaluate the performance of AMIC, we compare the execution times of AMIC against a brute force version on different data sizes. The brute force version computes the MI for every window without using incremental computation. To have a fair comparison, we only report the results in which the two algorithms run on one node. As shown in Table 3, our optimized version can achieve a speedup of 2, and the larger the data size, the larger the speedup. Note that the speedup will change depending on the data distribution. For example, if the data series are strongly correlated, the search will stop earlier, while if the data are weakly correlated, the search will take longer to identify significant windows.

TABLE 3: Execution times of optimized incremental algorithm and brute force algorithm

| #Samples | Brute Force | Incremental SW |
|---|---|---|
| 1K | 2.05 sec | 1.95 sec |
| 2K | 6.37 sec | 4.17 sec |
| 4K | 14.57 sec | 8.78 sec |
| 8K | 32.96 sec | 14.46 sec |
| 12K | 50.51 sec | 21.6 sec |
| 16K | 1 min 25.18 sec | 49.63 sec |
| 20K | 1 min 55.3 sec | 1 min 0.6 sec |
| 30K | 3 mins 1.44 sec | 1 min 27.04 sec |
| 50K | 8 mins 57.48 sec | 3 mins 39.04 sec |
| 100K | 21 mins 8.97 sec | 10 mins 9.05 sec |

*Scalability evaluation:* To evaluate the scalability of AMIC, we perform both stress test and scalability test on our Spark cluster, using synthetic data sets. The stress test is performed by fixing the number of executors used in the Spark cluster while changing the data sizes. In Figure 17a, the number of executors is fixed to 2 while the data size changes from 500 thousands to 2 millions data points. The stress test shows that AMIC can scale to large data sets with the execution time grows linearly to the data sizes. On the other hand, the scalability test is performed by fixing the data size while changing the number of executors. In Figure 17b, we use the data size containing 6 millions data points, and the number of executors ranging from 1 to 160 with 5 cores per executor. The scalability test shows that AMIC is well parallelized, and achieves significant speedup under parallelism (e.g., 6 millions data points are processed in less than 2 minutes).

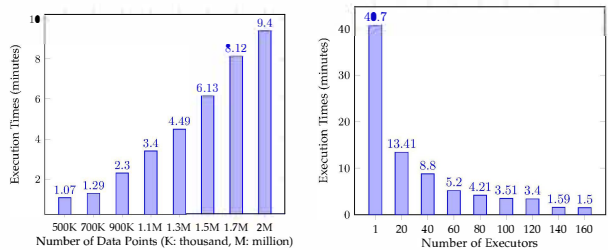

(a) Stress Test  (b) Scalability Test

Fig. 17: Stress Test and Scalability Test on Spark cluster

*Summary* In this section, we have performed an extensive evaluation on the performance of AMIC, verifying its capability in addressing Big Data challenges: variety, volume, velocity, and scalability. Specifically, the use of MI to measure correlations allows AMIC to uncover different types of relations and to work on any types of data, and thus to tackle the variety challenge. The layering approach with incremental streaming computation helps overcome the volume and velocity challenges. Finally, the use of Apache Spark ensures the scalability of the AMIC framework.

## 6 CONCLUSION

In this paper, we present AMIC, an adaptive and scalable method based on information-theoretical concept of mutual information to search for multi-scale temporal correlations in big data sets. The use of mutual information in searching for correlations helps tackle the variety challenge, whereas the hierarchical top-down search approach with incremental update helps overcome the volume and velocity challenges of Big Data. The method has been extensively evaluated using both synthetic and real-world data sets. With the presented results, we believe that the proposed method can be beneficial for both Data Science and Big Data communities, providing users an efficient tool to explore big and heterogeneous data sets.

The approach also suggests a new perspective on how we should deal with Big Data. In an era where data are massive, diverse and complex, a wise treatment for it is not to process entire data blindly, but to pre-select the most informative data partitions before performing any further analysis. By reducing the amount of data that are potentially not informative, different mining and learning algorithms can be beneficial in terms of computation efficiency and accuracy. For future extensions, several directions are promising. For example, AMIC can be extended to capture the spatial component of data, the time delayed correlation, or to discover the dependency across more than two variables.

## ACKNOWLEDGMENTS

This work has been partially supported by the DICYPS project funded by Innovation Fund Denmark, the GoFLEX project funded by the Horizon 2020 program, and the AAU International Postdoc Program.

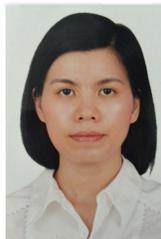

**Nguyen Ho** is a Postdoc Research Associate at the Center for Data-Intensive Systems (Daisy) at the Department of Computer Science, Aal-borg University, Denmark. Her research focuses on Big Data Analytics and Machine Learning methods for spatio-temporal multidimensional big data. Her current research interests include Big Data Analytics, Knowledge Extraction and Inference.

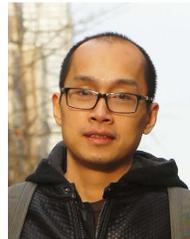

**Huy Vo** is an Assistant Professor of Computer Science at the City College of New York and a member of the doctoral faculty at the Graduate Center, City University of New York. He is also a faculty member at the Center for Urban Science and Progress, New York University. His current research focuses on high performance systems for interactive visualization and analysis of big data sets, specifically in urban applications. He has over ten years of research experience in large-scale data analysis and visualization, and co-authored over 40 technical papers and 3 patents, and contributed to several widely-used open-source systems.

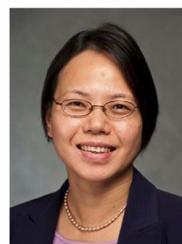

**Mai Vu** is an Associate Professor in Electrical Engineering at Tufts University, USA. She conducts research in wireless communication and networks, signal processing and information theory. Dr. Vu has served on the TPC of numerous IEEE conferences and during 2013-2016 served as an editor for the IEEE Transactions on Wireless Communications. She is a senior member of the IEEE.

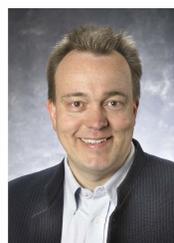

**Torben Bach Pedersen** is a Professor at the Center for Data-Intensive Systems (Daisy) at the Department of Computer Science, Aalborg University, Denmark. His research concerns business intelligence and big data, especially "Big Multidimensional Data" – The integration and analysis of large amounts of complex and highly dynamic multidimensional data. He is an ACM Distinguished scientist, a senior member of the IEEE, and a member of the Danish Academy of Technical Sciences.